\documentclass{article}

\PassOptionsToPackage{numbers}{natbib}
    \usepackage[preprint]{neurips_data_2022}
\usepackage[utf8]{inputenc} % allow utf-8 input
\usepackage[T1]{fontenc}    % use 8-bit T1 fonts
\usepackage{hyperref}       % hyperlinks
\usepackage{url}            % simple URL typesetting
\usepackage{booktabs}       % professional-quality tables
\usepackage{amsfonts}       % blackboard math symbols
\usepackage{nicefrac}       % compact symbols for 1/2, etc.
\usepackage{microtype}      % microtypography
\usepackage{xcolor}         % colors
\usepackage{bbm}
\usepackage{multirow}
\usepackage{graphicx}
\usepackage{enumitem}
\usepackage{wrapfig,lipsum,booktabs}
\usepackage{color, colortbl}
\usepackage[symbol]{footmisc}
\usepackage{float}

\definecolor{Gray}{gray}{0.95}

\usepackage[normalem]{ulem}
\hypersetup{
    colorlinks=true,
    }
\useunder{\uline}{\ul}{}

\title{Mr. Right: Multimodal Retrieval \\ on Representation of ImaGe witH Text}

\author{%
  Cheng-An Hsieh$^{1*}$, ~~
  Cheng-Ping Hsieh$^{2*}$, ~~
  Pu-Jen Cheng$^{1}$ \\
  $^1$National Taiwan University,~~
  $^2$UC San Diego \\
  ($^*$equal contribution) \\
  \texttt{ r09944010@ntu.edu.tw,~~ c2hsieh@ucsd.edu,~~ pjcheng@csie.ntu.edu.tw} \\
}

\begin{document}

\maketitle

\begin{abstract}
  Multimodal learning is a recent challenge that extends unimodal learning by generalizing its domain to diverse modalities, such as texts, images, or speech. This extension requires models to process and relate information from multiple modalities. In Information Retrieval, traditional retrieval tasks focus on the similarity between unimodal documents and queries, while image-text retrieval hypothesizes that most texts contain the scene context from images. This separation has ignored that real-world queries may involve text content, image captions, or both. To address this, we introduce Multimodal Retrieval on Representation of ImaGe witH Text (Mr. Right), a novel and comprehensive dataset for multimodal retrieval. We utilize the Wikipedia dataset with rich text-image examples and generate three types of text-based queries with different modality information: text-related, image-related, and mixed. To validate the effectiveness of our dataset, we provide a multimodal training paradigm and evaluate previous text retrieval and image retrieval frameworks.  The results show that proposed multimodal retrieval can improve retrieval performance, but creating a well-unified document representation with texts and images is still a challenge. We hope Mr. Right allows us to broaden current retrieval systems better and contributes to accelerating the advancement of multimodal learning in the Information Retrieval.
\end{abstract}
\section{Introduction}
Recent advancements in the field of digital media have resulted in a surge of interest in multimodal learning. Multimodal learning aims to learn well-unified representations from different modalities such as language, vision, or audio and projects them into a common low-dimensional space. For example, visual question answering needs an understanding of both vision and language \citep{antol2015vqa,goyal2017making,khattab2020colbert}; video highlight detection exploits video and audio features to identify the exciting moments \citep{videohightlight2021, videohightlight22021}; emotion recognition requires a fusion of spoken words, facial expressions, and voice \citep{busso2008iemocap, emotion-multimodal2018}.

In information retrieval, the conventional retrieval tasks focus on unimodal learning, including text-to-text \citep{nguyen2016ms,kwiatkowski2019natural} and image-to-image  \citep{philbin2007object,nister2006scalable,jegou2008hamming} retrieval. Both the texts and images contain comprehensive information, requiring the model to compute semantic representations of a single modality and match the unimodal document-query pair. Prior works \citep{khattab2020colbert,boytsov-nyberg-2020-flexible,zhang2021poolingformer,singh2021end,zheng2017sift,gordo2016deep,babenko2014neural} have improved retrieval performance and assisted users in searching for the requested documents. While beneficial, these tasks suffer from a significant key limitation: the text representations and the image features exist in their own spaces. In real-world applications, users may take texts or images as the queries to retrieve relevant data of the other modality. Therefore, cross-modal retrieval has attracted considerable attention from researchers recently.

% Further, with the development of digital media, many online materials and documents contain both texts and images, such as news, blogs, social media posts, and commercial websites. Therefore, we propose Mr. Right with text-image documents and generated text-based queries related to document texts, document images, or both. Further, we present a benchmark to evaluate how to balance an unified multimodal representation between texts and images.

% models have to encode image-text pairs into their respective features, and map them into a common low-dimensional space.

\begin{figure}
    \centering
    \includegraphics[width=0.9\textwidth]{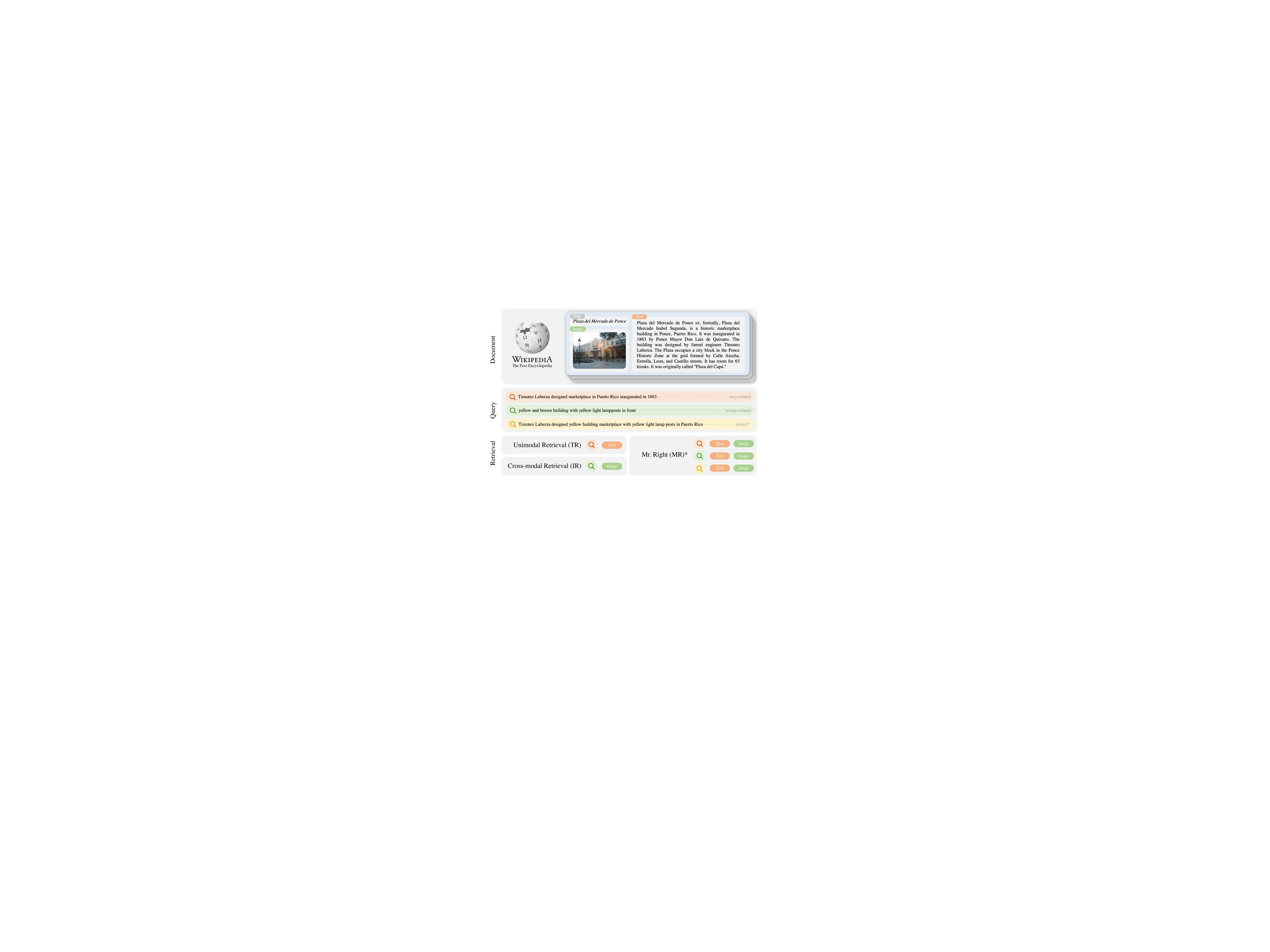}
    \vspace{-5pt}
    \caption{Overview of Mr. Right compared to unimodal and cross-modal retrieval. Unimodal retrieval (text-to-text retrieval) uses text-related queries to search for document texts, while cross-modal retrieval (text-to-image retrieval) uses image-related queries to find document images. Mr. Right utilizes each query and proposed mixed queries to retrieve documents with texts and images.}
    \label{fig:framework}
    \vspace{-15pt}
\end{figure}

Cross-modal retrieval aims to retrieve a relevant unimodal document from another modality of a query. Image-text retrieval is a fundamental challenge in cross-modal retrieval, and most existing methods \citep{dou2021empirical,li2021align,tan2019lxmert,chen2020uniter} train their models on the COCO \citep{lin2014microsoft} and Flickr30K \citep{plummer2015flickr30k} datasets. These datasets include images and their captions. However, unlike text-to-text retrieval, the texts in these datasets only have image related information. Nowadays, many online materials and documents may have both texts and images, such as Wikipedia, news, blogs, social media posts, and commercial websites. Also, considering most people utilize text-based queries to search multimedia documents on a search engine, a searching query can be keywords, captions, or both. The retrieval frameworks should be able to deal with the above text-based query with different modality information.

We introduce Mr. Right as shown in Figure~\ref{fig:framework}, a new comprehensive and challenging retrieval dataset, which provides text-image documents and text-based queries with different modality information that are text-related, image-related, or mixed. For documents, we collect from the Wikipedia-based Image Text Dataset \citep{srinivasan2021wit}, including paragraphs and photos. For queries, based on our knowledge, there is no existing dataset containing our proposed three types of queries. Therefore, we hire Amazon Mechanical Turk (AMT) workers to construct total 3k annotated queries for each type. Moreover, we also provide 350k auto-generated queries for model pre-training before learning annotated queries. In our training paradigm, we introduce document-query contrastive learning (DQC) and document-query matching (DQM) to fuse text and image features into multimodal representations.  

Finally, we propose a full-ranking benchmark for Mr. Right. Similar to TR datasets \citep{lin2014microsoft}, our full-ranking evaluation contains three types of annotated queries that models have to retrieve the most relevant document from all corpus. We create the benchmark based on our multimodal framework with the comparison to prior text-to-text retrieval (TR) and text-to-image retrieval (IR) baselines with human performance. Our results show that multimodal retrieval (MR) can perform better with the help of extended information from different modalities. Interestingly, it is a balance to incorporate these modalities into unified representations. However, it is still challenging to achieve comparable retrieval performance to the human benchmark. 

With Mr. Right, we take a significant step toward establishing a novel benchmark for evaluating the capabilities of multimodal retrieval systems. To the best of our knowledge, Mr. Right is the first multimodal retrieval dataset that explores multimodal documents and text-based queries with different modality information, and it is open-source to welcome methods of all kinds.

\section{Related Work}
\begin{table}[t]
\centering
\begin{center}
\resizebox{\textwidth}{!}{
\begin{tabular}{@{}lll|r|rr|rrr@{}}

\toprule
                    &                          &                 & \multicolumn{1}{c|}{\textbf{Train}}   & \multicolumn{2}{c|}{\textbf{Dev}} & \multicolumn{3}{c}{\textbf{Test}}                         \\
\textbf{Dataset}    & \textbf{Task}            & \textbf{Source} & \multicolumn{1}{c|}{\textbf{\#Pairs}} & \textbf{\#Pairs}    & \textbf{\#Query}    & \textbf{\#Pairs} & \textbf{\#Query} & \textbf{\#Document} \\ \toprule
MS MARCO \citep{nguyen2016ms}            & Text Retrieval           & Misc.           & 532,761          & ---                 & ---        & ---     & 6,980            & 8,841,823           \\
TREC-NEWS \citep{soboroff2018trec}           & Text Retrieval           & News            & ---              & ---                 & ---        & ---     & 57               & 594,977             \\
HOTPOTQA \citep{yang2018hotpotqa} & Question Answering & Wikipedia & 170,000          & ---                 & 5,447        & ---     & 7,405           & 5,233,329  \\
NQ \citep{kwiatkowski2019natural} & Question Answering & Wikipedia & 132,803 & --- & --- & --- & 3,452 &  2,681,468 \\
FEVER \citep{thorne2018fever} & Fact Checking & Wikipedia & 140,085 & --- & 6,666 & --- & 6,666 &  5,416,568 \\
\midrule
SOP \citep{oh2016deep} & Image-to-Image Retrieval & Products & 59,551 & --- & --- &  --- & 11,316 & 60,502 \\
CUB-200-2011 \citep{wah2011caltech} & Image-to-Image Retrieval & Birds &  5,864 & --- & --- & --- & 100 & 5,924 \\
$R$Oxford \citep{radenovic2018revisiting}             & Image-to-Image Retrieval & Landmark          & ---              & ---                 & ---        & ---     & 70               & 4,993                \\
Ukbench \citep{wang2011contextual}             & Image-to-Image Retrieval & Misc.           & ---              & ---                 & ---        & ---   & 2,550              & 10,200                 \\ \midrule
MS COCO \citep{lin2014microsoft}             & Image-Text Retrieval     & Misc.           & 118,000          & 5,000               & ---        & 5,000   & ---              & ---                 \\
Conceptual Captions \citep{sharma2018conceptual} & Image-Text Retrieval     & Misc.           & $\sim$3,300,000  & 2,800               & ---        & 2,300   & ---              & ---                 \\
Flickr30K \citep{plummer2015flickr30k}           & Image-Text Retrieval     & Flickr          & 30,000           & 1,000                & ---        & 1,000    & ---              & ---                 \\ 
\midrule
M5Product \citep{dong2022m5product}    & Mutlimodal Retrieval     & Products   & 4,423,160          & ---                 & ---      & ---     & 1,991            & 117,858 \\
Mr. Right (ours)    & Mutlimodal Retrieval     & Wikipedia       & 351,979 / 1,000          & ---                 & ---      & ---     & 2,047            & 806,357 \\
\bottomrule
\end{tabular}}
\vspace{5pt}
\caption{Statistics of datasets in retrieval tasks. Text-to-text retrieval datasets and image-to-image retrieval datasets contain queries and documents in a large search pool size. Image-text retrieval datasets include pairs of data. M5Product consists of  multi-modality products as documents and queries. Unlike M5Product, Mr. Right is the multimodal retrieval dataset exploring text-based query with different modality information. It involves around 35k auto-generated training pairs, 1k human-annotated pairs for fine-tuning, and 2k annotated pairs for testing. All pairs in our dataset are composed of one document and three types of queries.}
\label{tab:dataset}
\vspace{-10pt}
\end{center}
\end{table}

In this section, we describe previous retrieval datasets and explain how the existing methods employ transformer-based neural networks \citep{vaswani2017attention} to learn the representations of different modalities. Table~\ref{tab:dataset} shows an overview of the retrieval datasets.

\paragraph{Retrieval dataset} Most previous retrieval datasets only consider single modality (texts or images) documents without a unified representation among multiple domains. We categorize these datasets into unimodal and cross-modal learning. Text retrieval (TR) and image-to-image retrieval are the fundamental challenges in unimodal learning. Previous TR datasets \citep{nguyen2016ms,kwiatkowski2019natural,yang2018hotpotqa,soboroff2018trec,thorne2018fever} comprise a large corpus of text-based documents and related queries. They collect documents from different sources, such as Wikipedia \citep{yang2018hotpotqa,thorne2018fever}, news \citep{soboroff2018trec}, and online articles \citep{nguyen2016ms}. These sources involve diverse and generalized domain knowledge, reflecting real-world situations when users search from an extensive database. To ensure the quality of queries, some works collect the queries from searching logs \citep{nguyen2016ms,kwiatkowski2019natural} or hire crowd workers to generate annotations \citep{yang2018hotpotqa,thorne2018fever}. Similarly, the existing image-to-image datasets \citep{oh2016deep,wah2011caltech,radenovic2018revisiting,wang2011contextual} include several categories of images in the same domain, such as products, birds, and landmarks. Major works \citep{tan2021instance,ramzi2021robust,oh2016deep} randomly sample images from each category as queries, while the remaining images are the documents. As shown in Table~\ref{tab:dataset}, these unimodal datasets have more documents than queries, showing the challenge of ranking large numbers of documents. For cross-modal learning, the existing image-text retrieval (IR) datasets \citep{sharma2018conceptual,lin2014microsoft,plummer2015flickr30k} contain images and their captions. Major works harvest their images and captions from the web. They develop a pipeline to extract, filter, and transform their captions. In this task, the number of images equals the captions, meaning documents and queries are in pairs.  Unlike the unimodal tasks having a large size of documents, the evaluation of cross-modal performs on a small size of document-query pairs. Different from single modality documents, recent work \citep{m5product} has proposed an E-commerce product multimodal retrieval dataset that contains data more than two modalities.

% models have to measure how similar a given pair of images according to the category-level or object-level.

% The documents are collected from different sources, such as Wikipedia, News, and online articles. 

% Previous generated queries including QA (cite HotSpotQA) and fact checking (cite FEVER)

\paragraph{Retrieval model}
Due to the superior performance of contextualized representations in transformer-based models \citep{vaswani2017attention}, self-attention-based architectures have become the model of choice in natural language processing (NLP) and computer vision (CV). In unimodal retrieval, the existing methods \citep{santhanam2021colbertv2,karpukhin2020dense,xiong2020approximate,qu2020rocketqa} of text-to-text retrieval employ transformers to encode queries and documents into vector representations and compute their similarity. Also, the Vision Transformers \citep{dosovitskiy2020image} reduce the time-consuming process of extracting region features, and the later works \citep{el2021training,li2022hashformer,chen2021transhash} attain excellent results in image-to-image retrieval. In cross-modal retrieval, Vision-and-Language Pre-training (VLP) models have improved performance on IR tasks. The recent CLIP
\citep{radford2021learning} and ALIGN \citep{jia2021scaling} utilize contrastive learning to align the unimodal representations of image-text pairs. Other VLP methods (\emph{e.g.}\ METER \citep{dou2021empirical}, ALBEF \citep{li2021align}, LXMERT \citep{tan2019lxmert}, UNITER \citep{chen2020uniter}) perform multimodal fusion to produce the joint representations of text-image pairs. This bridges the semantic gap between visual and textual features in texts and images.

\section{The Mr. Right Dataset}
\label{sec-dataset}
\begin{figure}
    \centering
    \includegraphics[width=\textwidth]{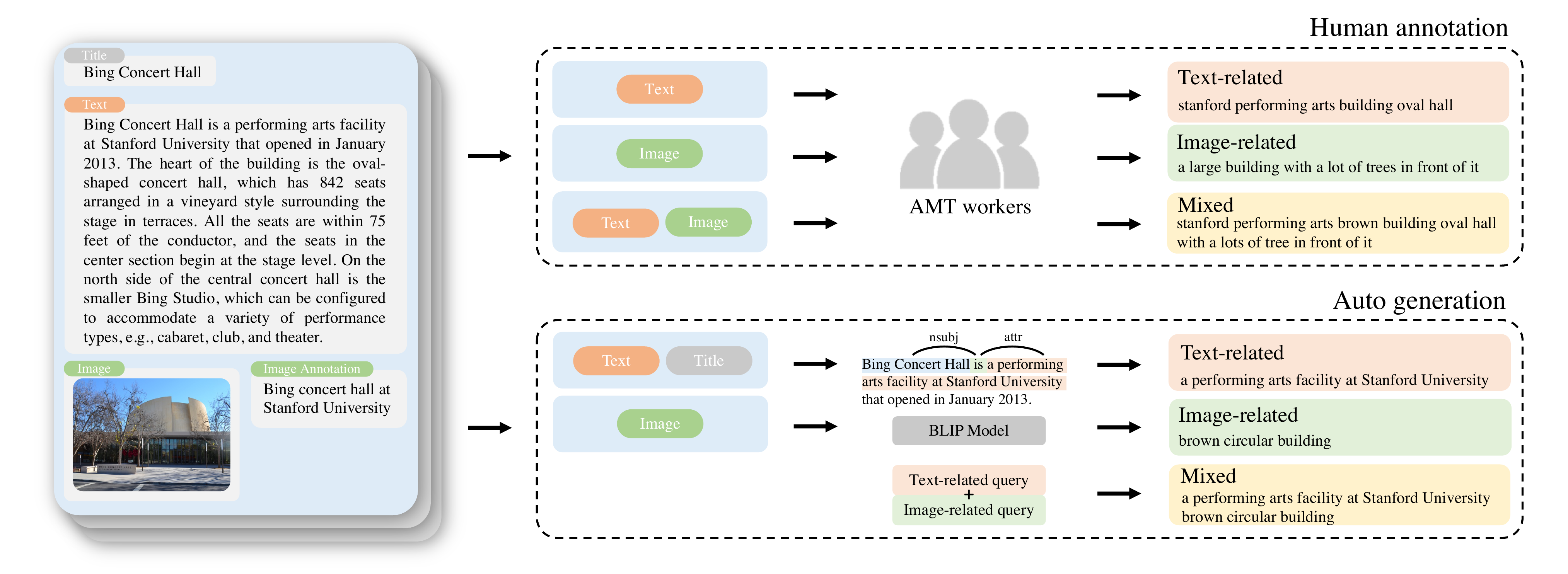}
    \caption{Overview of Mr. Right's data collection process. Human-annotated queries are created by AMT workers following our annotation instructions, while auto-generated queries are constructed with dependency parsing, image captioning, and concatenation.}
    \label{fig:dataset}
\end{figure}

Mr. Right aims to construct a new dataset for multimodal retrieval tasks. The dataset focuses on two components: (1) Multimodal documents consist of different modality information, including texts and images; (2) Text-based queries involve text content, image captions, or both. Mr. Right collects documents and annotated/generated queries by extracting, labeling, and filtering.

% Retrieval systems should be able to deal with different types of queries simultaneously. 
% We develop an automated pipeline to generate queries for training, and compare with human-annotated queries.

\subsection{Data Collection}
\paragraph{Wikipedia-based document}
To generate multimodal documents with diverse knowledge domains, we gather paragraphs and photos from the Wikipedia-based Image Text Dataset \citep{srinivasan2021wit}, which consists of 37.6 million entity-rich image-text pairs with 11.5 million unique images. The original dataset includes 108 languages, and we only keep 1.5 million English data for simplification. We process this dataset with three steps: (1) image filtering eliminates images with invalid download links, corrupted content, and non-JPEG/PNG images; (2) text filtering removes repeated pages and deletes contents without mentioning the title, ensuring that the document is related to the subject; (3) text reduction extracts the first paragraph as the document to avoid high memory and computational requirements. The whole Wikipedia article can be very long, and we find that usually the first paragraph contains a brief introduction. After the filtering, there are 806,357 of multimodal documents remaining. % 64.42\%
% text reduction extracts the first paragraph with a brief introduction as the document to save the memory and computational requirements.

% We tokenize the document text based on BERT model and discover that up to 26.65\% of documents are over 128 tokens. This may increase the memory and computational requirements due to the self-attention operation on Transformer-based models. \cp{not need mention self-attention} To address this, we analyze the text content, and find that most of the wiki passages have an introduction in the first paragraph. Therefore, we extract the first paragraph as the document text. In total, the average document length is 46
% % , and the average document tokens is 
% \ca{where to introduce the title, document, image? maybe a figure?} \cp{The figure in second page will have a example document title/text/image and the difference of unimodal/crossmodal/multimodal retrieval} \cp{I think this paragraph is too detailed...You should write with in the previous paragraph with only 2-3 sentences. You can just say the whole wiki document will have what problems? And we use the first paragraph because of what?}

\paragraph{Human-annotated query}
A query may relate to the document text, document image, or both. To generate three types of queries based on multimodal documents, we hire crowd-workers with qualifications from Amazon’s Mechanical Turk (see Appendix~\ref{data-collection}). The annotators are shown a text, a image, and both successively to come up with queries based on first impression. To ensure the consistency and quality of annotations, we give the following guidelines: (1) Word count limitation. Each query should be between 10 to 100 words to ensure the query involves enough information. (2) No title. Annotators should avoid including the title because it is the result that users want to retrieve. (3) No copied phrases in the passage. A real-world query may involve ambiguous meaning or terms, so it is better to paraphrase the sentences. (4) Requirement of adjectives and nouns in the image query. Instead of only phrasing the objects in the image, we hope crowd-workers describe the details, such as colors, shapes, and actions. The annotated query examples can be found in Figure~\ref{fig:dataset}.

\paragraph{Auto-generated query}
Annotating queries for the whole multimodal dataset can be time-consuming. To address this, we develop a pipeline that extracts snippets of the document texts as the text-related queries and generates captions of the document images as the image-related queries. According to Figure \ref{fig:dataset}, Wikipedia content generally has a specified format. The first sentence begins with the title and a short introduction, followed by the details. Therefore, we utilize dependency parsing using spaCy API and detect the dependent verb of the title in the first sentence. Then we take the adjectives and nouns after the verb as the text-related query and remove the adjectives in the first sentence as the document. This ensures that models have to learn the information from the remaining texts. For the image-related queries, each image in the original Wikipedia dataset has its own annotation. However, many annotations relate to the document title instead of the image content. In addition, some of the annotations contain proper names, which is difficult to learn from the scene context. To address this, we replace them with generated image captions based on BLIP \citep{li2022blip} that outperforms a variety of methods on vision and language tasks. As shown in Figure~\ref{fig:dataset}, the caption generated from the model is closer to the image content, and it can be the query that human uses to search for visual information. To generate the mixed queries, we concatenate the former two queries, which contain text information and image context respectively.

% for an automated and efficient approach to index, categorize
% and organize visual media with little to no human intervention
% is growing in multiple fields like medicine??
% we construct our dataset with the following methodology.

\subsection{Annotated Query Validation}
% To ensure the quality of human-annotated queries, we design two filtering method.

\paragraph{Rule-based filtering} A well-defined query should have multiple part-of-speech (POS) tags. Therefore, annotated query candidates without nouns or with only one noun lacking adjectives or verbs are discarded. Since queries directly copied from the documents are trivial for retrieval, we drop text-related and mixed query candidates that highly overlap with document texts. We remove the queries with the longest-common-substring (LCS) length larger than 40 and the ratio (divided by query length) larger than 0.6. As for image-related queries, we take out the candidates that include additional knowledge more than the document image context. In other words, we filter out the queries containing proper names, such as a particular person's identity or locations. The above three filters discard around 10\% of the candidates.

\paragraph{Human filtering}
In our task, each query should correspond to a unique document. To ensure uniqueness, we utilize text retrieval model BM25 and image retrieval model CLIP to search relevant documents given text-related and image-related queries, respectively. For simplicity, we retrieve top-10 relevant candidates from the whole multimodal documents and prioritize to examine the queries without unique document pairs, i.e., close ranking scores for different documents. After filtering out these queries, we efficiently validate whether the semantic meaning between the query and the correct document is unrepeated. After our validation, there are 25\% of query sets discarded and remain 3,047 annotated query sets.

After finishing the collection and validation stage, our dataset contains 806,357 multimodal documents, 351,979 auto-generated query sets, and 3,047 human-annotated query sets. Each query set is mapped to one document and contains three types of proposed queries. We further split 1k human-annotated sets for fine-tuning and 2k for testing as shown in Table~\ref{tab:dataset}.

\begin{table}[t]
\centering
\resizebox{\textwidth}{!}{
\begin{tabular}{rccccc}
\toprule
& \textbf{SECS} & \textbf{} & \textbf{Vocab. Size} & \textbf{Avg. Word Lengths} & \textbf{Top-3 NER Tag}                \\ \midrule

\rowcolor{Gray} \multicolumn{5}{l}{\textit{Human-annotated}} \\ 
Text-related        & 3,047            & 8,290                & 9.54                       & GPE (32.14\%), DATE (14.62\%), NORP (10.70\%)     \\ 
Image-related       & 3,047            & 2,609                & 8.26                       & CARDINAL (59.49\%), NORP (24.65\%), PERSON (3.68\%) \\
Mixed              & 3,047            & 8,752                & 13.58                      & GPE (31.66\%), DATE (13.49\%), NORP (11.19\%) \\  

\rowcolor{Gray} \multicolumn{5}{l}{\textit{Auto-generated}} \\ 
Text-related        & 351,979            & 55,928               &  4.95                      & NORP (41.54\%), GPE (24.82\%), ORG (8.77\%)     \\ 
Image-related       & 351,979            & 13,464                & 9.83                       & CARDINAL (51.58 \%), GPE (34.62 \%), NORP (4.79\%)  \\
Mixed              & 351,979            & 62,440                & 14.78                      & NORP (39.41\%), GPE (25.48\%), ORG (7.90\%) \\

\bottomrule
\end{tabular}}
\vspace{2pt}
\caption{Analysis of annotated and generated queries in Mr. Right. Top-3 NER tags show the top entities classified in the whole corpus. GPE means geopolitical entities, NORP represents affiliations, and CARDINAL are numerical values.}
\label{tab:data-analysis}
\vspace{-5pt}
\end{table}

\begin{table}[t]
\centering
\resizebox{\textwidth}{!}{
\begin{tabular}{llccp{11cm}}
\toprule 
\textbf{Query type}  & \textbf{Properties} & \textbf{A\%} & \textbf{G\%} & \textbf{Example(s)}                                                                          \\ \midrule
Text-related         & \cellcolor{Gray}Paraphrase          & \cellcolor{Gray}53                  & \cellcolor{Gray}0                    & \cellcolor{Gray} \textbf{A: } Seoul based artist, animator famous for eyedolls$^{\dag}$                                       \\
                     & Keyword Extraction  & 23                  & 0                    & \textbf{A: } Railway station in Schwelm in Rhine Westaphalia$^{\ddag}$                                        \\
                     & \cellcolor{Gray}Duplication         & \cellcolor{Gray}24                  & \cellcolor{Gray}100                  & \cellcolor{Gray} \textbf{A: } Northern Alabama pygmy sunfish$^{\ast}$                                                         \\
                     & \cellcolor{Gray}                      &  \cellcolor{Gray}                   &    \cellcolor{Gray}                  & \cellcolor{Gray} \textbf{G: } A species of pygmy sunfish$^{\ast}$                                                                   \\ \midrule
Image-related        &   One object          & 28                  & 19                   & \textbf{A: } Small swimming gray and white fish$^{\ast}$                                                     \\
                     &                     &                     &                      & \textbf{G: } fish that is swimming$^{\ast}$                                                                        \\
                     & \cellcolor{Gray}Multiple objects    & \cellcolor{Gray}72                  & \cellcolor{Gray}81                   & \cellcolor{Gray} \textbf{A: } Girl with straight brown and pink hair in white dress on white background$^{\dag}$              \\
                     & \cellcolor{Gray}                    &   \cellcolor{Gray}                  &  \cellcolor{Gray}                    & \cellcolor{Gray} \textbf{G: } A woman with pink hair and a white shirt$^{\dag}$                                                    \\ \midrule
Mixed                &   Fusion              & 47                  & 0                    & \textbf{A: } A German railway station which runs through thick green forest$^{\ddag}$                         \\
                     & \cellcolor{Gray}Concatenation       & \cellcolor{Gray}53                  & \cellcolor{Gray}100                  & \cellcolor{Gray} \textbf{A: } Small swimming gray and white fish Northern Alabama pygmy sunfish$^{\ast}$                      \\
                     & \cellcolor{Gray}                    &   \cellcolor{Gray}                  &  \cellcolor{Gray}                     & \cellcolor{Gray} \textbf{G: } A species of pygmy sunfish a fish that is swimming in some water$^{\ast}$ \\     
\bottomrule                               
\end{tabular}}
\vspace{2pt}
\caption{Properties distribution of annotated queries (A) and generated queries (G). Each query type has 100 samples, and we manually categorize each query into different properties. We find annotated queries have more diverse properties than generated ones. The corresponding documents of provided examples are listed in Appendix~\ref{sec:usage}}
\label{tab:data-category}
\vspace{-5pt}
\end{table}

\subsection{Dataset Analysis}
\paragraph{Quantitatively analysis}
Table~\ref{tab:data-analysis} presents statistics for our annotated and generated queries showing vocab sizes, average lengths, and top-3 named entity recognition (NER) tags. Each type of query has the same amount. Based on the vocabulary size, we can find that text-related queries have more diverse words than image-related queries, indicating that our image descriptions are composed of limited illustrative words. Furthermore, texts contain more sparse information than images, resulting in the need for longer word lengths of annotated text-related queries. The top-3 NER tags suggest that human favors distinct entities such as countries and dates, while our generated approach mainly focuses on affiliations in the text-related queries. For image-related queries, human and the BLIP model prefer to describe the number of objects.

% that annotators leverage more different words to create text-related queries. This is because they can refer to the text document and try to paraphrase the sentence. 

\paragraph{Qualitatively analysis}
We heuristically identify query properties covered in the dataset to recognize the difference between human-annotated and auto-generated queries. We randomly sample 100 queries from the three types of queries and present the results in Table~\ref{tab:data-category}. As can be seen, we split the properties of text-related queries into three categories. Paraphrase means that queries involve different words and sentence clause structure from the documents; keyword extraction indicates that queries only include important terms; duplication means queries are the reorganized document phrases. Considering efficiency, we copy the snippets of document texts to generate text-related queries. For the image-related queries, some annotators may focus on describing the most conspicuous object, while others include multiple objects and their adjectives. Auto-generation produces more image-related queries with multiple objects. It may be because the BLIP model learns captioning from many data and prefers to describe the image details. The difference between annotated and generated mixed queries is also apparent. Human may fuse the image descriptions and text content or concatenate them with prepositions. Our generated mixed queries only rely on the concatenation. In our study, annotated queries contain more diverse types of queries and well-unified sentence structures, but the annotation process is time-consuming and expensive. Our auto-generated queries ensure efficiency, and the experiment results show that these queries are effective.

% NER Tags
% Statistics Length, unique tokens...
% Relation between human query and pseudo query (similarity scores?)
% Sample 100 query to categorize different types

\subsection{Benchmark} 
% Follow Wild (real-world) 950k
To simulate the real-world retrieval problems, we create Mr. Right's benchmark with the whole corpus of 800k Wikipedia documents. This full-ranking setting guarantees that the model can handle large numbers of multimodal documents. Further, search queries may have text keywords or image descriptions, so we present three retrieval tasks with the corresponding queries: text-related, image-related, and mixed. See Appendix~\ref{appendix-benchmark}. for more task details.

% In order to deal with the real-world retrieval problem, our multimodal retrieval dataset can get close to the real documents with both texts and images while queries are like real search situations containing keywords and descriptions in different modality. To simulate true user search patterns, we present three main retrieval tasks based on different queries in the followings:

\subsection{Evaluation Metrics}
Retrieval tasks might be precision-focused or recall-focused, depending on the requirements of real-world applications. In Mr. Right, documents and queries are binary relevant, and retrieving relevant documents from our large corpus of different modalities is challenging. Following previous image-text retrieval tasks \citep{dou2021empirical,li2021align,tan2019lxmert,chen2020uniter}, we report recall@$k$ as our performance metric. Further, considering the rank of documents, we also utilize MRR (Mean Reciprocal Rate) as our binary rank-aware metric, a general standard in text retrieval tasks. In our experiments, we compute recall with $k=1,5,10$ and MRR@10 for all models and assess their performance.
\section{Multimodal Retrieval}
\label{sec-model}
With Mr. Right, the next step is to set up the retrieval task based on the multimodal documents and text-based queries. We illustrate our retrieval formulation (Section \ref{Retrieval Formulation}) and model architecture (Section \ref{Model Architecture}). Then we describe our two training objectives (Section \ref{Training Objectives}).

\subsection{Retrieval Formulation}
\label{Retrieval Formulation}
Given a document $D$ with a paragraph text $D_{T}$ and an image $D_{I}$, we use a multimodal encoder to fuse $D_{T}$ and $D_{I}$ into a single fixed-size multimodal vector representation $R_{d}$. Also, we encode a text-based query $Q_{T}$ into a fixed-size vector representation with our text encoder. To establish our retrieval task, we need to encode all the documents $\{(D^1_{T}, D^1_{I}), (D^2_{T}, D^2_{I}),...,(D^N_{T}, D^N_{I})\}$ and queries $\{Q^1_{T},Q^2_{T},...,Q^M_{T}\}$ into $\{R^1_{d},R^2_{d},...,R^N_{d}\}$ and $\{R^1_{q},R^2_{q},...,R^M_{q}\}$ respectively. With these representations, we compute the cosine similarity scores between documents and queries and find the most similar document for each query. In this scenario, we can build offline indexing for document representations and compute query representations online for real-world applications.

\subsection{Model Architecture}
\label{Model Architecture}
\paragraph{Document (Multimodal) Encoder}
To encode both document texts and images into unified multimodal representations, we leverage previous pre-trained VLP models for initialization in our framework. These models have learned a common low-dimensional space to embed vision and language features. In these models, we have a vision encoder (\emph{e.g.} CNNs or vision transformers \citep{dosovitskiy2020image}) and a text encoder (\emph{e.g.} BERT \citep{devlin2018bert} or RoBERTa \citep{liu2019roberta}) to extract modality-specific features. Then we have a fusion module (\emph{e.g.} co-attention or merge-attention \cite{dou2021empirical}) to integrate both features into a unified feature. Therefore, we can view these VLP models like a black box multimodal encoder $E_M$ to output size-variant multimodal document features $F_d$ whose size depends on the length of input texts $D_T$ and the dimension of images $D_I$. To derive a single fixed-size representation $R_d$ for each document, we simply average the size-variant document features $F_d$.

\paragraph{Query (Text) Encoder}
Since our queries are text-based $Q_T$, we create a query encoder $E_q$ and share the parameters from the text encoder of our multimodal encoder. This ensures the text representations of queries are similar to document texts. Like a multimodal encoder, we take the average of the query features $F_q$ to obtain query representations $R_q$.

\subsection{Training Objectives}
\label{Training Objectives}
In this section, we introduce document-query contrastive learning (DQC) and document-query matching (DQM) to project document and query representations into the same space. 
% Moreover, we make use of context prediction to ensure the multimodal representation to contain both document text $D_T$ and query text $Q_T$ information.

\paragraph{Document-Query Contrastive learning}
Contrastive learning has been widely used to train on VLP models \cite{radford2021learning,jia2021scaling,li2021align} which can increase the similarity scores between parallel pairs. With document and query representation $R_d$ and $R_q$, we learn two projection functions $f_d$ and $f_q$ with a fully-connected layer to map their representations into the same space. Then we calculate the cosine similarities between document and query pairs in a training batch. The matched pairs are positive while all other pairs are negative. Based on the pairs, we minimize the contrastive loss $L_{dqc}$ like in-batch cross-entropy loss. To organize in-batch negatives more effectively, we keep two queues \cite{li2021align} to store the most recent $K$ representations, enlarging the amounts of negative samples per batch.

\paragraph{Document-Query Matching}
To further learn a fine-grained similarity of pair documents and queries, we build a binary classifier $C$ to predict whether the output features of document encoder $F_d$ and query encoder $F_q$ is matched. Specifically, we use a 6-layer transformer model and insert a special token $[CLS]$ at the head of the input sequence to obtain global information. Then we employ a linear classifier on this token followed by softmax to predict a two-class label and compute the matching loss $L_{dqm}$ according to binary cross-entropy loss. Motivated by ALBEF \cite{li2021align}, we sample online hard negative pairs for each document and query from contrastive similarity distribution. In addition to real-world negative documents, we produce two pseudo negative documents by combining the positive document and the sampled negative document into pairs of a positive image and a negative text and vice versa. Hence, for a query, we have a positive document, a sampled negative document, and two pseudo negative documents.

% \paragraph{Context Prediction}
% It is possible that our multimodal (document) representation will only remain partial information and discard original details. Thus, we propose context prediction to reconstruct the representation with both bag of document words and query words. Actually, we reuse the text encoder input embeddings $Emb$ to linear project the single fixed-size representation to a vocabulary size $V$ vector. Our label $y_{cp}$ is a soft one-hot bag of words vector with one indicating the presence of the word token while zero denoting the absence. After, we normalized with the total counts of word tokens for each label. The whole context prediction loss is: 
% \begin{equation}
%     y^{ij}_{cp} = \frac{\mathbbm{1}_{BoW(D^i_T, Q^i_T)}(j)}{\sum^V_{k=1}\mathbbm{1}_{BoW(D^i_T, Q^i_T)}(k)},\;\; \hat{y}^{ij}_{cp} = \frac{\exp(Emb_j(R^i_d))}{\sum^V_{k=1}\exp(Emb_k(R^i_d))}
% \end{equation}
% \begin{equation}
% L_{CP}=-\frac{1}{B}\sum^{B}_{i} \sum^{V}_{j}y^{ij}_{cp}\log(\hat{y}^{ij}_{cp})
% \end{equation}

\section{Experiments}
\label{sec-exp}

\subsection{Dataset} 
Mr. Right has both auto-generated and human-annotated queries. We first pre-train our models on the 350k auto-generated document-query pairs to learn the multimodal representations. Further, we fine-tune our learned model on the human-annotated 1k training pairs with 10\% as our validation set.

\subsection{Baselines}
We compare our proposed multimodal retrieval framework with TR and IR baselines. We only collect existing dense retrieval \citep{denseretrieval2018} approaches for a fair comparison. Additionally, we develop the MR baseline with the ensemble of TR and IR baselines. Text retrieval models only consider the document texts; image retrieval models only focus on the document images; multimodal retrieval models perceive both document texts and images.  All the baseline models are described in Appendix~\ref{appendix-baselines}.

\subsection{Experiment Setup}
\label{Experimental Setup}
We train our framework using existing VLP models, including METER \citep{dou2021empirical}, ALBEF \citep{li2021align}, and ViLT \citep{kim2021vilt} to make use of their multimodal pre-trained weights. The pre-training process lasts for 40 epochs and fine-tunes for 20 epochs on 8 NVIDIA V100 GPUs. Our optimizer is AdamW with a weight decay of 0.02, and the learning rate is warmed-up to $5 \times 10^{-5}$ in the first epoch and decayed to $1 \times 10^{-7}$ following the scheduler. Also, we set up a 0.5 gradient clipping value and 9,600 queue size for DQC. For image augmentation, we use random-crop of size 288$\times$288 or 384$\times$384 depending on the pre-trained VLP models and apply RandAugment \citep{cubuk2020randaugment}. For texts, we truncate our max length for queries with 40 and documents with 128. In order to simulate real-world user queries, we randomly select text-related, image-related, or mixed queries during training. 

\begin{table*}[t]
\centering
\begin{center}
\resizebox{\textwidth}{!}{
\begin{tabular}{cr|rrrr|rrrr|rrrr}

\toprule
& \multirow{2}{*}{Method} & \multicolumn{4}{c|}{text-related query} & \multicolumn{4}{c|}{image-related query} & 
\multicolumn{4}{c}{mixed query} \\
& & MRR@10 & R@1 & R@5 & R@10 & MRR@10 & R@1 & R@5 & R@10 & MRR@10 & R@1 & R@5 & R@10 \\
\midrule
% & BM25 & 54.5 & 71.4 & 76.8 & 61.9 & 0.3 & 1.1 & 1.4 & 0.7 & 45.4 & 62.8 & 68.0 & 53.0\\

\multirow{6}{*}{TR} & RoBERTa (ZS) & 0.0 & 0.0 & 0.1 & 0.2 & 0.0 & 0.0 & 0.0 & 0.0 & 0.0 & 0.0 & 0.0 & 0.0 \\
& DiffCSE (ZS) & 32.0 & 24.5 & 42.6 & 49.8 & 0.2 & 0.1 & 0.3 & 0.4 & 20.2 & 14.9 & 27.9 & 32.7 \\
& SBERT (ZS) & 35.6 & 27.3 & 46.5 & 54.4 & 1.0 & 0.5 & 1.5 & 2.2 & 28.2 & 21.4 & 37.3 & 43.9 \\

& RoBERTa (FT) & 25.5 & 18.8 & 34.6 & 41.9 & 0.4 & 0.2 & 0.9 & 1.5 & 24.0 & 17.1 & 33.9 & 40.7 \\
& DiffCSE (FT) & 33.3 & 26.3 & 42.8 & 51.0 & 0.6 & 0.3 & 0.9 & 1.7 & 30.1 & 22.8 & 40.2 & 47.7 \\
& SBERT (FT)   & 47.7 & 38.7 & 60.0 & 68.0 & 1.1 & 0.5 & 1.9 & 2.6 & 39.3 & 31.0 & 50.2 & 58.8 \\

\midrule

\multirow{4}{*}{IR} & CLIP (ZS) & 2.6 & 1.6 & 3.7 & 5.7 & 5.1 & 3.3 & 7.4 & 10.2 & 7.0 & 4.1 & 10.8 & 14.7 \\
& ALBEF (ZS) & 0.5 & 0.3 & 0.6 & 1.2 & 3.7 & 2.2 & 5.6 & 8.0 & 1.8 & 0.8 & 2.9 & 4.5 \\
& CLIP (FT) & 3.3 & 1.9 & 5.0 & 7.5 & 5.7 & 3.6 & 8.0 & 11.9 & 8.3 & 4.8 & 12.7 & 18.6 \\
& ALBEF (FT) & 0.9 & 0.5 & 1.5 & 2.1 & 6.2 & 4.0 & 9.3 & 12.9 & 4.1 & 2.1 & 6.5 & 9.4 \\

\midrule

% & BM25 + ALBEF (FT) & 54.5 & 71.4 & 76.8 & 61.9 & 4.6 & 10.5 & 13.9 & 7.1 & 55.9 & 72.8 & 77.4 & 63.2 \\
\multirow{4}{*}{MR} & SBERT (FT) + ALBEF (FT) & \textbf{48.0} & \textbf{39.0} & 60.7 & 67.9  & 7.2 & 4.4 & 10.9 & 15.3  & 50.3 & \textbf{41.6} & 61.8 & 70.3 \\
& Our (METER) &  38.8 & 27.4 & 54.8 & 64.3 & 12.0 & 7.2 & 18.3 & 25.3  & 42.9 & 31.6 & 58.2 & 69.2 \\
& Our (ALBEF) &  44.4 & 32.3 & \textbf{61.1} & \textbf{72.4}  & 4.2 & 2.1 & 6.2 & 10.8 & \textbf{50.7} & 38.0 & \textbf{67.2} & \textbf{78.2}\\
& Our (ViLT) &  26.8 & 16.4 & 40.6 & 53.4 & \textbf{15.7} & \textbf{8.6} & \textbf{23.0} & \textbf{33.0} & 45.4 & 33.0 & 62.7 & 73.5 \\

\bottomrule

\end{tabular}}
\caption{Retrieval performance across three benchmark tasks with different types of queries. We compare MRR and recall@$k$ among baselines and our proposed models on TR, IR and MR. ZS is zero-shot, and FT is fine-tuned.}
\label{tab:main}
\vspace{-10pt}
\end{center}
\end{table*}

\subsection{Results and Analysis}

\paragraph{Compared to TR/IR} 
We present the retrieval results in Table~\ref{tab:main}. We compare our method against TR/IR models and discuss the performance difference across three query types. The table shows that TR and IR have difficulties responding to the opposite queries. TR obtains worse results for image-related queries while IR is vice versa. This may be because their documents only contain unimodal information with either texts or images. In contrast, MR shows the ability to mitigate this problem. It achieves comparable performance as TR on text-related queries and scores higher than IR on image-related queries. This improvement indicates that MR can perform better due to the extended information from different modalities. Further, when queries are mixed, MR exploits the advantage of multimodal representations and achieves superior performance compared to TR and IR.

% \paragraph{Compared to BM25} \label{exp-bm25} \ca{remove?} However, a big challenge is to compare neural-based models with statistics-based method BM25 which achieving the best results for both text-related and mixed query in our table. Since the queries in our dataset are largely composed of document keywords or duplicated phrases, it is beneficial to use BM25 by collecting the bag-of-words. On the other hand, neural-based approaches aim to encode the semantic of whole documents or queries rather than the focus on specific words. Although BM25 outperforms neural-based methods, it can hardly handle image-related query.

% \input{Image/fig-attention}

\paragraph{Multimodal representation} \label{exp-mr}  We integrate fine-tuned SBERT and ALBEF as an ensemble MR model that shows a comparable performance to our MR models. Although incorporating TR and IR models can perform well among different types of queries, the vector size of document representations for multiple modalities increases linearly, and we need to define the best weighted combination of their output scores. In contrast, our proposed multimodal representation can unify multiple domain information into a standard size feature. To understand the multimodal representations, we compute Grad-CAM visualizations (see Appendix~\ref{appendix-gradcam}) on the attention maps of document texts and images given different types of queries. The attention heat is highly correlated to where human would look to match the corresponding query. In Table~\ref{tab:main}, we find a trade-off of our framework to deal with text-related and image-related queries simultaneously. Comparing MR with different backbone VLMs, the performance is debated between the two queries. This may come from the limited size of our unified representation. We cannot include all the document text and image information together but a balance between them. 

\begin{wraptable}{r}{0.3\textwidth}
\centering
\vspace{-10pt}
\resizebox{0.3\textwidth}{!}{
\begin{tabular}{r|c}
\toprule
          & Accuracy \% \\
\midrule
% BM25       & 70.7     \\
Our (METER) & 30.0     \\
Our (ALBEF) & 26.0     \\
Our (ViLT)  & 26.7     \\
\midrule
Human      & 89.3     \\
\bottomrule
\end{tabular}}
\caption{Human evaluation on 150 random sampled queries.}
\label{tab:human}
\vspace{-10pt}
\end{wraptable}
\paragraph{Human evaluation}
Besides model performance, we also present human evaluation results compared to our MR models in Table~\ref{tab:human}. We randomly select 50 samples for each query type. To efficiently retrieve related documents for human evaluation, we utilize our MR model (METER) to obtain the top 3 relevant candidates with the correct document and construct a four-choice question with one correct answer. Table~\ref{tab:human} shows humans get 89.3\% accuracy. It validates the reliability of our dataset, but it also shows there is room for improvement of models on Mr. Right. It may be because human can understand various query properties in Table~\ref{tab:data-category}, extract the crucial text content, or perceive image scene context in detail. To see the retrieval results difference between our models and human, we provide failed examples in Appendix~\ref{appendix-mistake}. Also we provide the performance comparison of our auto-generated and human-annotated queries in Appendix~\ref{appendix-queries}.

% \subsection{Analysis}
% \paragraph{Grad-CAM visualizations} 

% \paragraph{Misaligned Examples}
% TBD.
% Figure for BM25 correct / Our model wrong / human wrong

\section{Conclusion}
In this paper, we propose Mr. Right, a multimodal retrieval dataset for information retrieval. Mr. Right covers three types of text-based search queries with different modality information, including text-related, image-related, and mixed, to simulate real-world search situations. Further, our dataset provides documents with texts and images to develop multimodal representation. We build our end-to-end multimodal retrieval model for Mr. Right to unify features across modalities. Compared to the previous text and image retrieval frameworks, multimodal retrieval shows improvements on different queries and points out the balance between modalities. However, current multimodal models still have a significant gap to human performance, showing the potential of Mr. Right as a challenge in multimodal retrieval. We believe Mr. Right can breathe new insights into information retrieval for more robust retrieval systems.

\section{Limitations and Future Work}
In Mr. Right, we only consider text-based queries, which may limit the search modalities from users. We can expand our dataset with additional domain queries and documents such as images, audio, and video. Further, Mr. Right focuses on the materials in Wikipedia. We can explore other sources such as news, blogs, or commercial websites. Mr. Right is a preliminary attempt to explore multimodal retrieval, and there are still challenges we need to analyze and study in future work.

\bibliographystyle{unsrt}
\bibliography{main} % add your ref to ref.bib
\newpage
\newpage
\appendix

% Offensive content
% License from other dataset

\section{Supplementary Materials for Mr. Right}
\label{sec:A}
We provide the following detailed sections and materials that complement the discussions in the main paper. Code and dataset are available at \url{https://github.com/hsiehjackson/Mr.Right}

\begin{itemize}[leftmargin=0.35cm]
\item Establishment of the datasheet for Mr. Right in Appendix~\ref{appendix-datasheet}

\item Details of the evaluation process in Appendix~\ref{appendix-evaluation}.

\item Designs of the proposed model in Appendix~\ref{appendix-model}.

\item Confirmations of the data license in  Appendix~\ref{appendix-license}.

\item Maintenance of Mr. Right in Appendix~\ref{appendix-maintenance}. 

\end{itemize}

% \appendix
\section{Datasheets}
\label{appendix-datasheet}
\subsection{Motivation}
Information retrieval is a fundamental and essential challenge in real-world applications. In the past, researchers focused on unimodal retrieval because previous datasets only included data with a single modality, such as text-to-text and image-to-image retrieval datasets. They design robust and effective frameworks to improve the performance of these retrieval tasks. However, humans perceive the world with different modalities, such as language, vision, or audio. Due to multimedia development, humans have begun to utilize one modality to search for another modality. For example, image-text retrieval is a challenge in which models need to learn a common representation between images and texts and retrieve the most relevant documents. Further, sometimes we may need to combine different modalities and understand the meaning together. To accelerate the advancement of retrieval on multimodal learning, we propose Mr. Right, which contains multimodal documents and three types of text-based queries according to the real-world context. It has 806,357 multimodal documents, 351,979 auto-generated queries, and 3,047 human-annotated queries for each type.

\subsection{Collection Process}
\label{data-collection}
\paragraph{Multimodal document}
We construct Mr. Right based on the Wikipedia-based Image Text (WIT) Dataset \citep{srinivasan2021wit}. The original dataset includes Wikipedia articles and Wikipedia image links in 108 languages. Each article has a page title, a page description, and a reference image description. The dataset has filtered the image-text pairs based on effective restrictions, such as text length, image size, and image format. However, Wikipedia updates its content frequently, some image URLs are outdated, and some pages have different versions. Therefore, we create our pipeline to filter WIT and obtain the multimodal documents. The process is explained in the following:

\begin{itemize}[leftmargin=0.35cm]
\item Download Wikipedia CSV file \citep{srinivasan2021wit} and keep English articles with titles in the content. There are about 1,479,330 English documents. Download the images using the Python \textit{multiprocessing} and \textit{urllib2} module. During the downloading, we find that some image URLs are invalid. It may be because Wikipedia has updated the links. Corrupted images are also discarded. After the downloading, there are 953,042 images that occupy 1.5TB.

\item Discard the documents with the same title. We analyze the composition of the remaining document candidates and find that some documents present the same title with the similar content. It is because the page may be updated according to the time, and there are different versions of the documents. To avoid one query mapping to multiple correct documents, we filter these repeated documents. Finally, we obtained 806,357 multimodal documents, including text-image pairs with rich semantic information.

\end{itemize}

\begin{figure}
    \centering
    \includegraphics[width=0.9\textwidth]{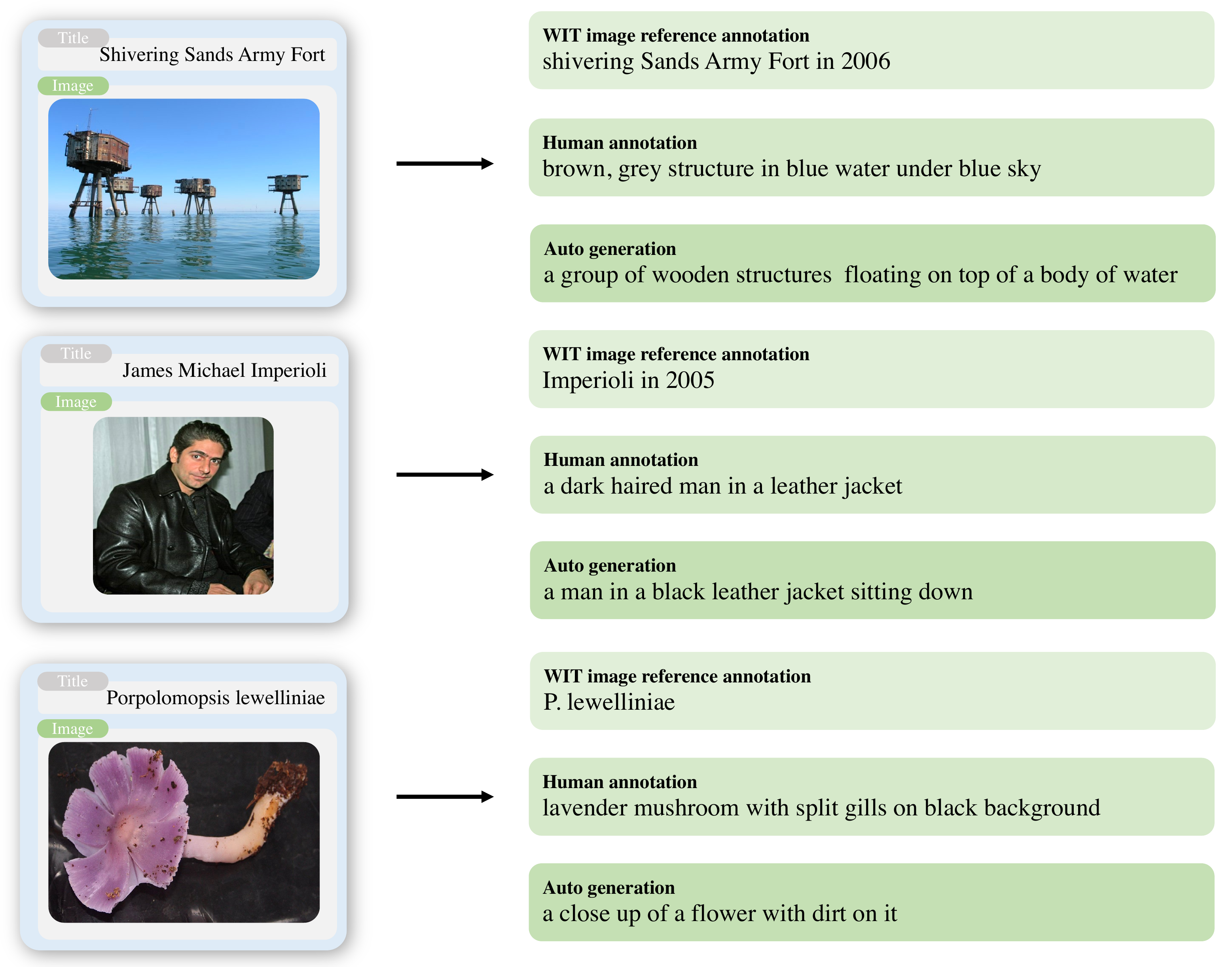}
    \vspace{-5pt}
    \caption{Examples of image-related query. }
    \vspace{-5pt}
    \label{fig:datasheetannotation}
\end{figure}

\begin{figure}
    \centering
    \includegraphics[width=\textwidth]{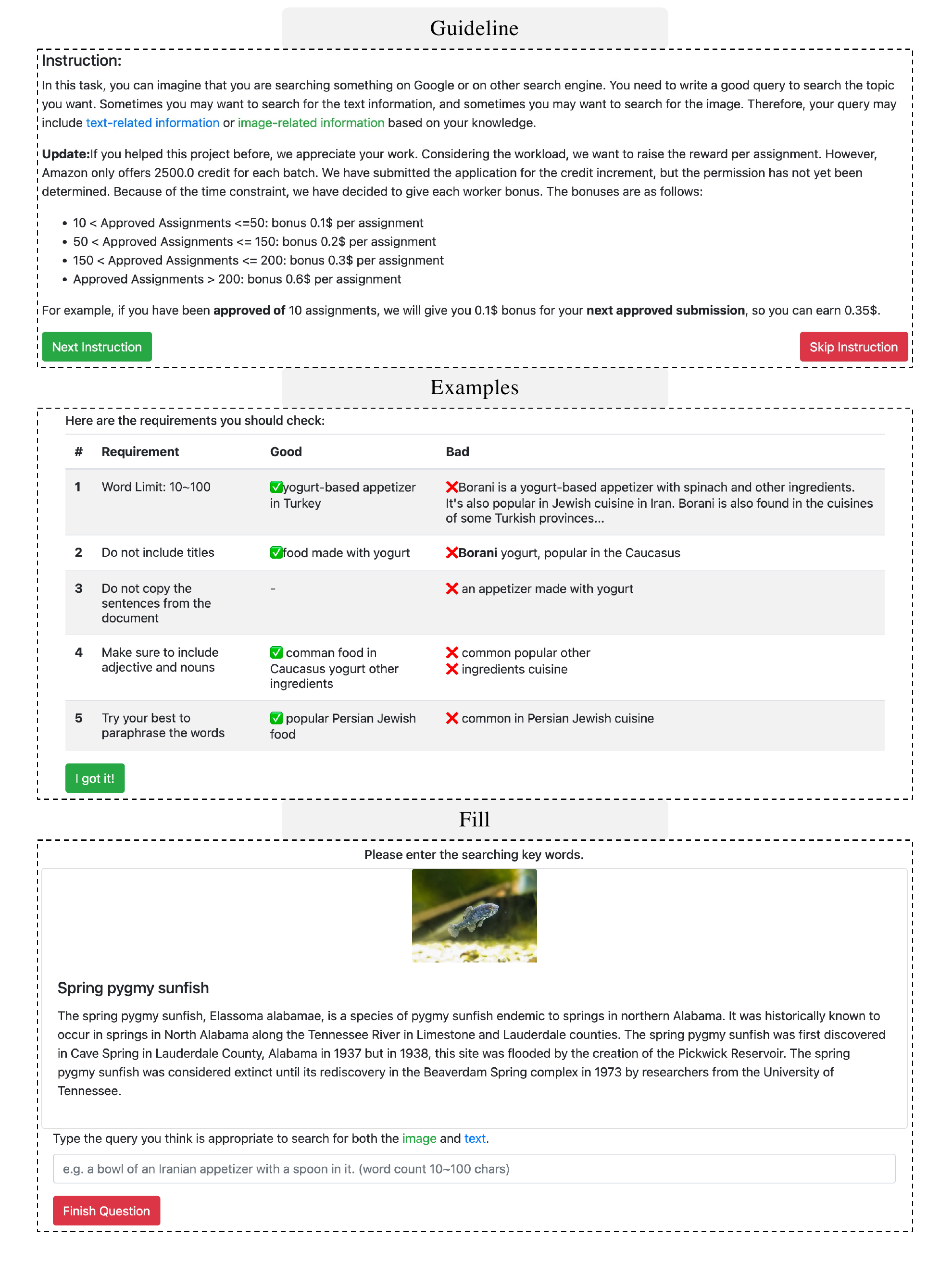}
    \caption{Template of human annotation. Annotators read the guideline and examples first, and then create the queries.}
    \label{fig:template}
\end{figure}

\newpage
\paragraph{Human-annotated query}
% Screeshot 
As shown in Figure \ref{fig:datasheetannotation}, WIT original image reference annotations contain page title or page description rather than the image context. In real-world applications, we consider that user queries may be image descriptions that include image objects, colors, background, or people's actions. Further, user queries may involve multimodal information, such as image caption fused with text content. In our study, there is no dataset that consists of mixed query for retrieval. Therefore, we hire annotators from Amazon Mechanical Turk to produce human-annotated queries. We require annotators should be masters to ensure label quality. Only annotator with at least 50 approved HITs and an 80\% HIT approval rate is allowed. We pay 0.25\$ USD per assignment that includes text-related query, image-related query, and mixed. Also, to award those hardworking annotators, we provide an additional bonus. After the annotation, the statistical data shows that  workers' average time per assignment (three types of queries) is 6 minutes 34 seconds. More details can be seen in Figure \ref{fig:template}. In total, we have paid 3,687.24\$ USD (including the platform fees) to annotate 4,276 assignments.

To further ensure the quality of Mr. Right, we provide guidelines and examples to human annotators. They have to read the guidelines first before labeling. Guidelines indicate that a query should meet some restrictions to simulate the possible real-world searching queries, and annotators can come up with the queries based on their habits by following the guidelines. The annotation template is illustrated in Figure \ref{fig:template},  and the guidelines are described as follows:

\begin{itemize}[leftmargin=0.35cm]
\item Words Limit: 10 -- 100 
\item Do not include title.
\item Do not copy the sentence from the document.
\item Try your best to paraphrase the words.
\item Include image information such as color, gender, action, etc.
\item Include adjectives and nouns for images.
\end{itemize}

\paragraph{Auto-generated query}
% Process time 
% Load which model
% Use which package
Coming up with searching queries is time-consuming. Therefore, we propose auto-generation for training queries. For text-related queries, we extract a snippet from the first sentence. Our analysis finds that most Wiki passages start with a page title and a brief introduction. Therefore, we utilize this format to extract the sentence's crucial information. We use Spacy API with en\_core\_web\_lg package to parse the sentence and detect the title-dependent verb. Then we take the snippet after the verb as the query. To increase the robustness of models, we also remove the snippet from the document text, which means models have to learn the representation from the remaining text and still be capable of matching the document-query pairs. For image-related queries, we implement the BLIP \citep{li2022blip} model, which outperforms many VLP frameworks on image captioning. We employ the model on 351,979 images and produce one caption for each image. The images are resized to 384×384. We use beam search with a beam size of 3 and set the maximum generation length as 30. For the mixed query,  it is still challenging to produce a query that fuses the text and image information. Considering the efficiency, we concatenate the text and image queries as mixed queries.

\begin{figure}[t]
    \centering
    \includegraphics[width=\textwidth]{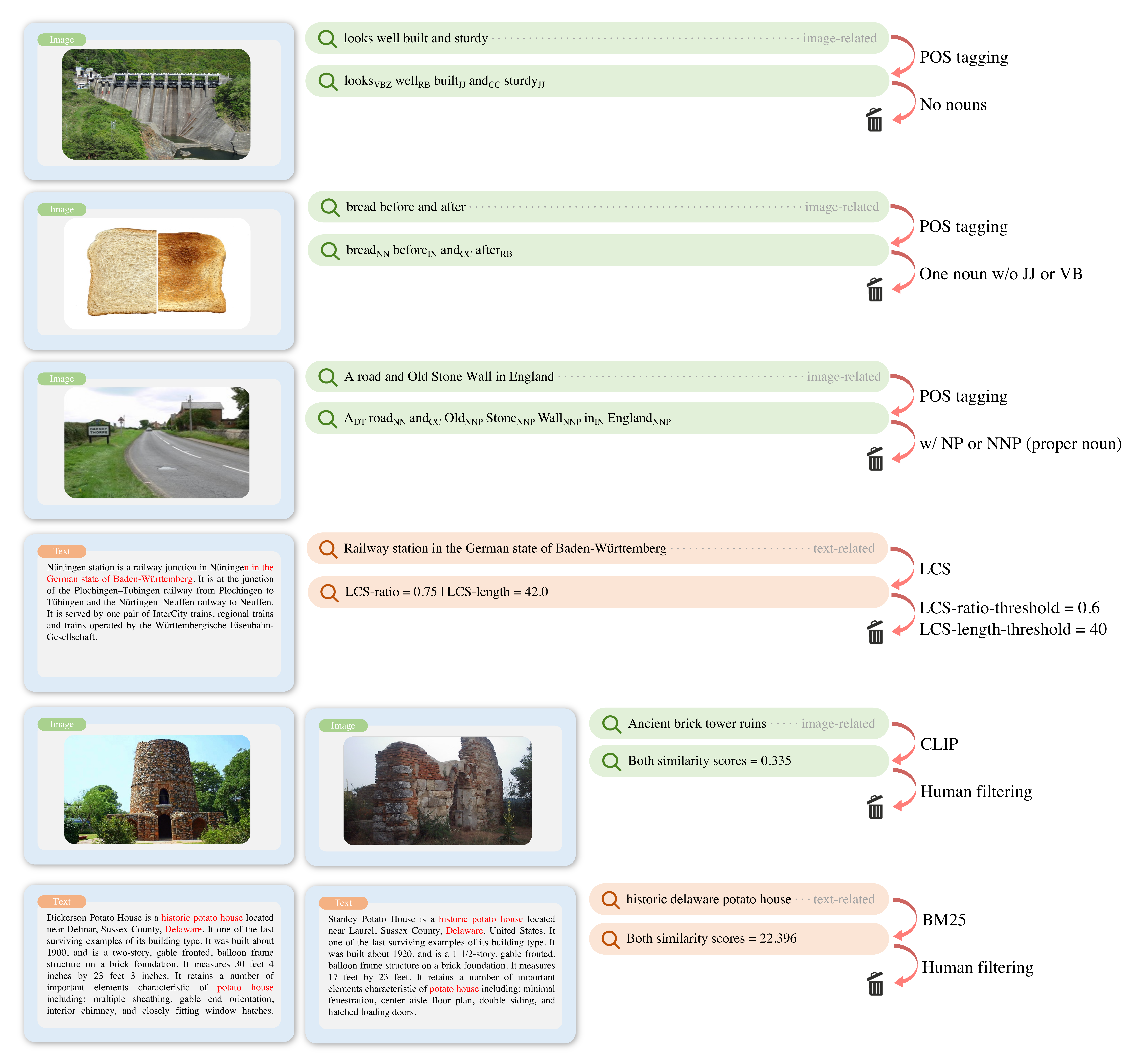}
    \vspace{-15pt}
    \caption{Process of annotated query validation  with the help of POS tagging, LCS, CLIP, and BM25.}
    \vspace{-10pt}
    \label{fig:filter}
\end{figure}

%  the words ``army''  and ``tank'' will attend corresponding parts on image while the words ``Australia'' and ``War'' will highlight ``Australian'' and ``fightning'' on texts
\subsection{Filtering}
In Figure~\ref{fig:filter}, we show the examples of annotated query validation, including rule-based and human filtering. For the first three examples, We filter out the queries through POS tagging. For the fourth example, we drop the queries by calculating the LCS. For the last two examples, we use CLIP and BM25 to support human discarding queries which map to ambiguous documents.

\begin{figure*}[h]
    \centering
    \includegraphics[width=\textwidth]{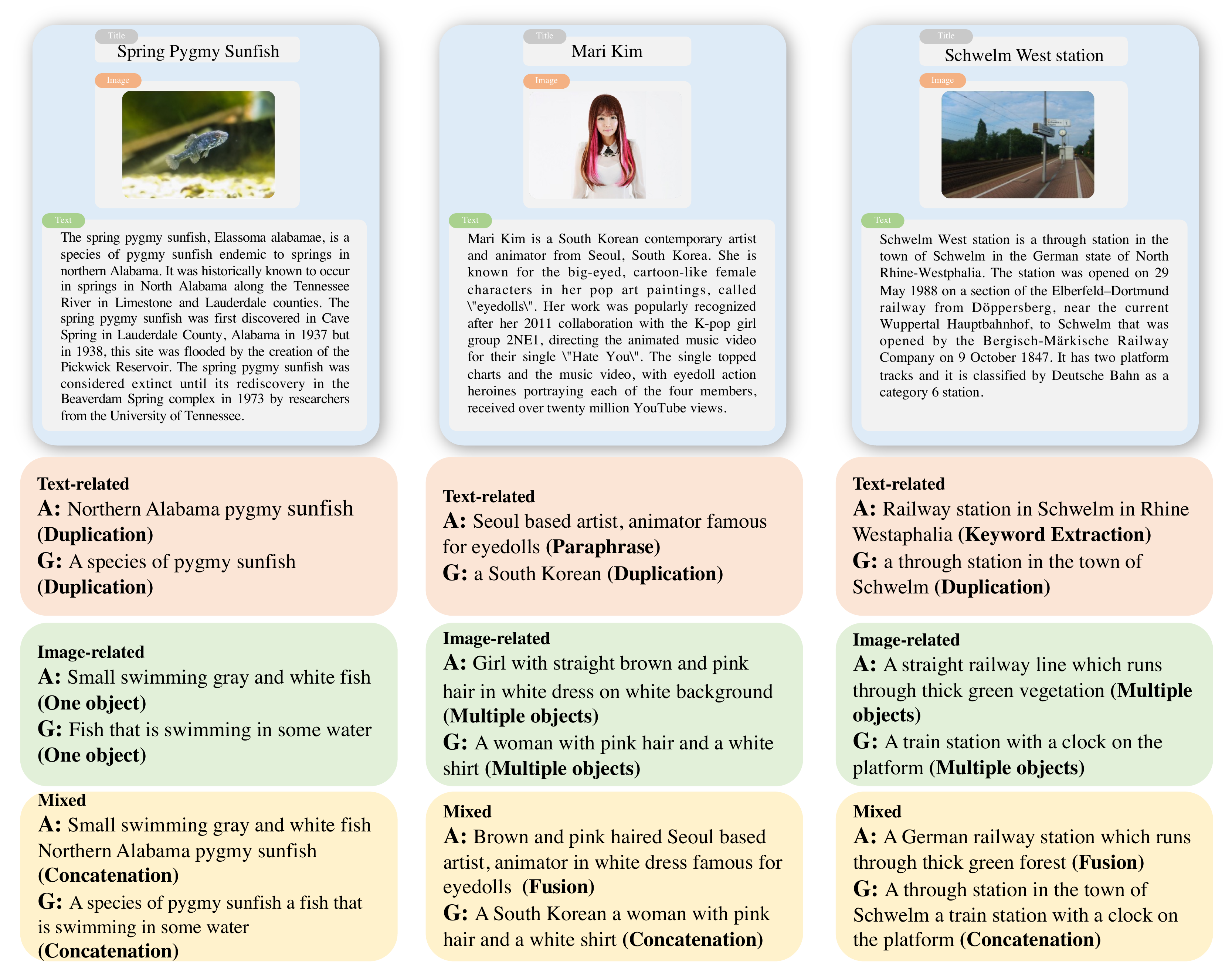}
    \caption{Examples of document and query pair with the corresponding property. The left document$^{\ast}$ has text-related query by duplication, image-related query with one object, and concatenation-based mixed query. The middle document${\dag}$ has text-related annotated query by paraphrasing and image-rated query with multiple objects. The right document${\ddag}$ has text-related annotated query by keyword extraction and mixed annotated query by fusion.}
    \label{fig:analysis-category}
\end{figure*}
\subsection{Usage}
We split Mr. Right into five files: \textit{multimodal\_documents.json}, \textit{multimodal\_pretrain\_pairs.json}, \textit{multimodal\_finetune\_pairs.json}, \textit{multimodal\_val\_queries.json}, and \textit{multimodal\_test\_queries.json}. In \textit{multimodal\_documents.json}, it contains document ids, titles, texts, and image URLs. We do not provide image files directly due to the copyright issue. In  \textit{multimodal\_pretrain\_pairs.json}, we provide our auto-generated queries and edited document texts. We still equip this file with the original document texts to keep the flexibility of using Mr. Right. Researchers can create their model framework and train on our auto-generated document-query pairs or produce other effective data. 
In \textit{multimodal\_finetune\_queries.json}, we randomly sample human-annotated document pairs for fine-tuning. In \textit{multimodal\_val\_queries.json} and \textit{multimodal\_test\_queries.json}, they include corresponding document ids and human-annotated queries. The examples of multimodal document-query pairs are shown in Figure~\ref{fig:analysis-category}. All of our source codes are uploaded to GitHub. Researchers can download json files from our repository. We also offer our training codes.
\label{sec:usage}
\newpage
\section{Evaluation Details}
\label{appendix-evaluation}
%  \cp{where is human evaluation website?}
\subsection{Benchmark tasks}
\label{appendix-benchmark}
In this section, we provide the benchmarks of Mr. Right on humans, baseline retrieval models, and our multimodal framework. There are three types of tasks, and the details are as follows:
\paragraph{Task1: Text-related query} This task aims to follow previous text retrieval datasets \citep{nguyen2016ms,kwiatkowski2019natural,yang2018hotpotqa,soboroff2018trec,thorne2018fever}. Users mostly search for documents relying on the keywords from document paragraphs or text-based information. In our dataset, text-related queries contain name entities (person, date, organization, location) or factual knowledge (relations, terminologies). 

\paragraph{Task2: Image-related query} This task aims to follow previous text-to-image retrieval datasets \citep{sharma2018conceptual,lin2014microsoft,plummer2015flickr30k}. With a blurred impression about the appearance of an object, users search documents based on the part of context from document images. In our dataset, image-related queries are similar to image captions that explain details of objects, such as color, shape, amount, position, or action.

\paragraph{Task3: Mixed query} We propose this task to simulate users searching documents with text-related and image-related information. To precisely find the correct document, Mr. Right provides document texts and images to consider both modalities for retrieval. Our mixed queries generate a brief description that includes the document paragraph and photo, which can be viewed as a combination of corresponding text-related and image-related queries.

\subsection{Baseline models}
\label{appendix-baselines}
\paragraph{Text retrieval models} 
% 1) BM25 \citep{robertson2009probabilistic}: a commonly-used bag-of-words retrieval function based on token-matching between two high-dimensional sparse vectors with TF-IDF token weights. 
To evaluate text retrieval performance with state-of-the-art (SOTA) neural frameworks, we test three approaches in the followings. 1) RoBERTa-base \citep{liu2019roberta}: a pre-trained language model which can encode both documents and queries into the contextualized sentence representations to compute the similarity in the same vector space. 2) DiffCSE \citep{chuang2022diffcse}: current unsupervised SOTA among sentence representation learning methods. 3) all-mpnet-base-v2 (SBERT): current supervised SOTA for sentence embedding tasks and semantic search tasks on SentenceTransformers \citep{reimers-2019-sentence-bert} leaderboard. We evaluate both zero-shot and fine-tuned performance for these models. We train with an in-batch negative loss function and use an AdamW optimizer with learning rate $2 \times 10^{-5}$ and batch size 32 for 30 epochs.

\paragraph{Image retrieval models}
Current image-text retrieval models are highly related to pre-trained vision-and-language models. Training with natural language supervision, these models demonstrate the ability of crossmodal image retrieval. We zero-shot evaluate CLIP \citep{radford2021learning}, and ALBEF \citep{li2021align} as our baselines and also fine-tune on our dataset. We don't use other VLP models (e.g., METER \citep{dou2021empirical}, ViLT \citep{kim2021vilt}) because most of them require a high computational overhead for evaluation due to the need to calculate matching scores across all image text pairs. We fine-tune CLIP and ALBEF using an Adam optimizer with learning rate $1 \times 10^{-6}$ and batch size 128 for 40 epochs.

\paragraph{Multimodal retrieval models}
Without existing baselines, we build an ensemble model by integrating the document-query similarity scores from our best TR model and IR model. We fuse the scores with a weighted sum parameter tuning on the validation set for different tasks. The final ensemble relevance scores are then used to rank the search results.  

\begin{figure}[H]
    \centering
    \includegraphics[width=\textwidth]{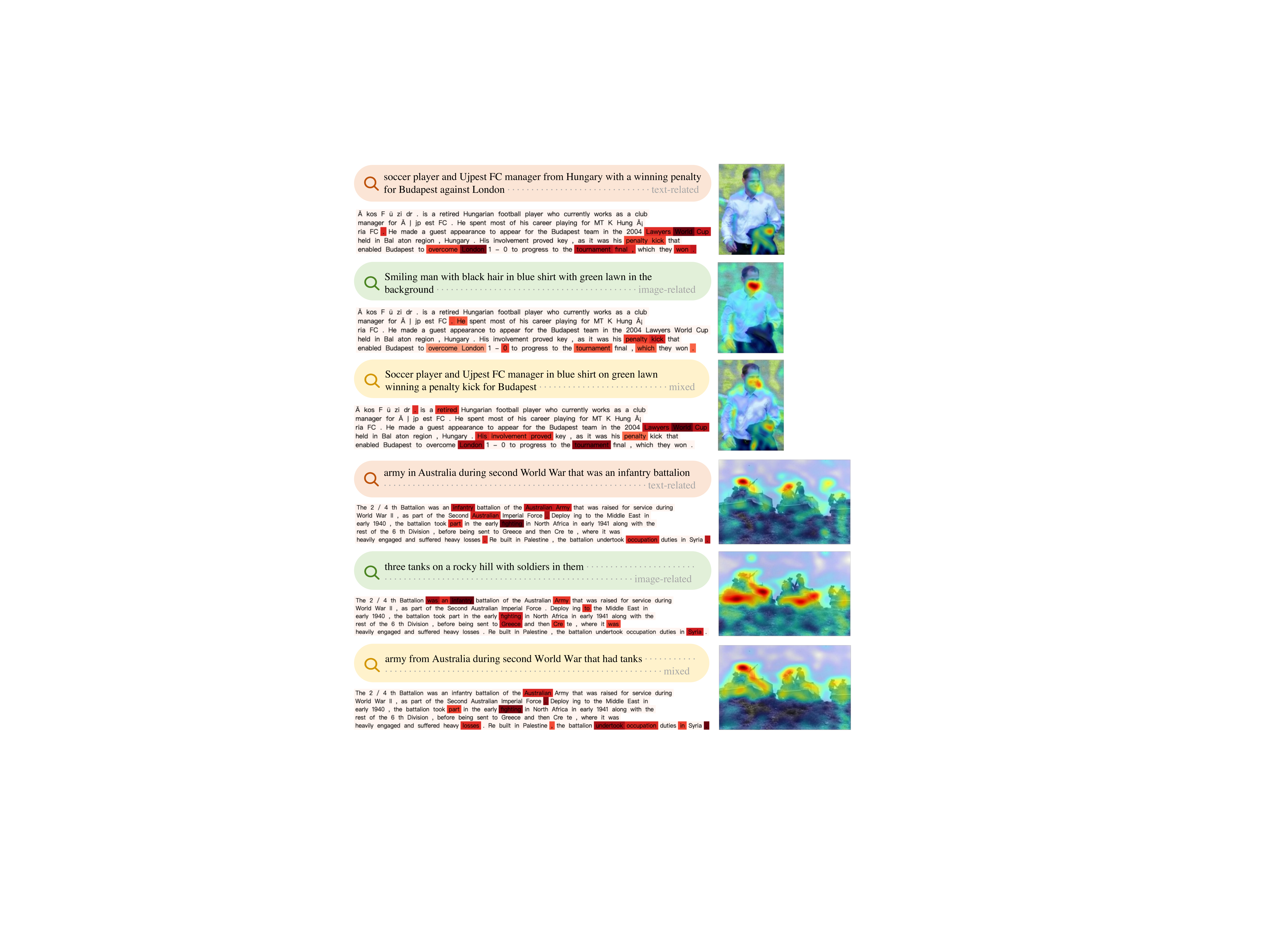}
    \caption{Grad-CAM visualizations on the cross-attention maps of the query-document matching classifier corresponding to different queries. Heats are scattered across document image and texts.}
    \label{fig:attention}
    % \vspace{-15pt}
\end{figure}

%  the words ``army''  and ``tank'' will attend corresponding parts on image while the words ``Australia'' and ``War'' will highlight ``Australian'' and ``fightning'' on texts
\subsection{Grad-CAM visualizations}
To better understand the multimodal representations of documents, we compute Grad-CAM visualizations on the cross-attention maps of the query-document matching classifier in Figure~\ref{fig:attention}. With different queries, our model will interact with different parts of the image and texts, which is highly correlated to where humans would look to match the pairs. For the above example, the word ``smiling'' highly focus on the face of the image, and the word ``winning'' is related to ``tournament'' and ``Lawyers World Cup'' in texts. For the bottom example, the words ``army'' and ``tank'' in the queries will attend to corresponding parts of the image, while the words ``Australian'' and ``War'' will highlight ``Australian'' and ``fighting'' in texts, respectively.
\label{appendix-gradcam}

\begin{figure}
    \centering
    \includegraphics[width=\textwidth]{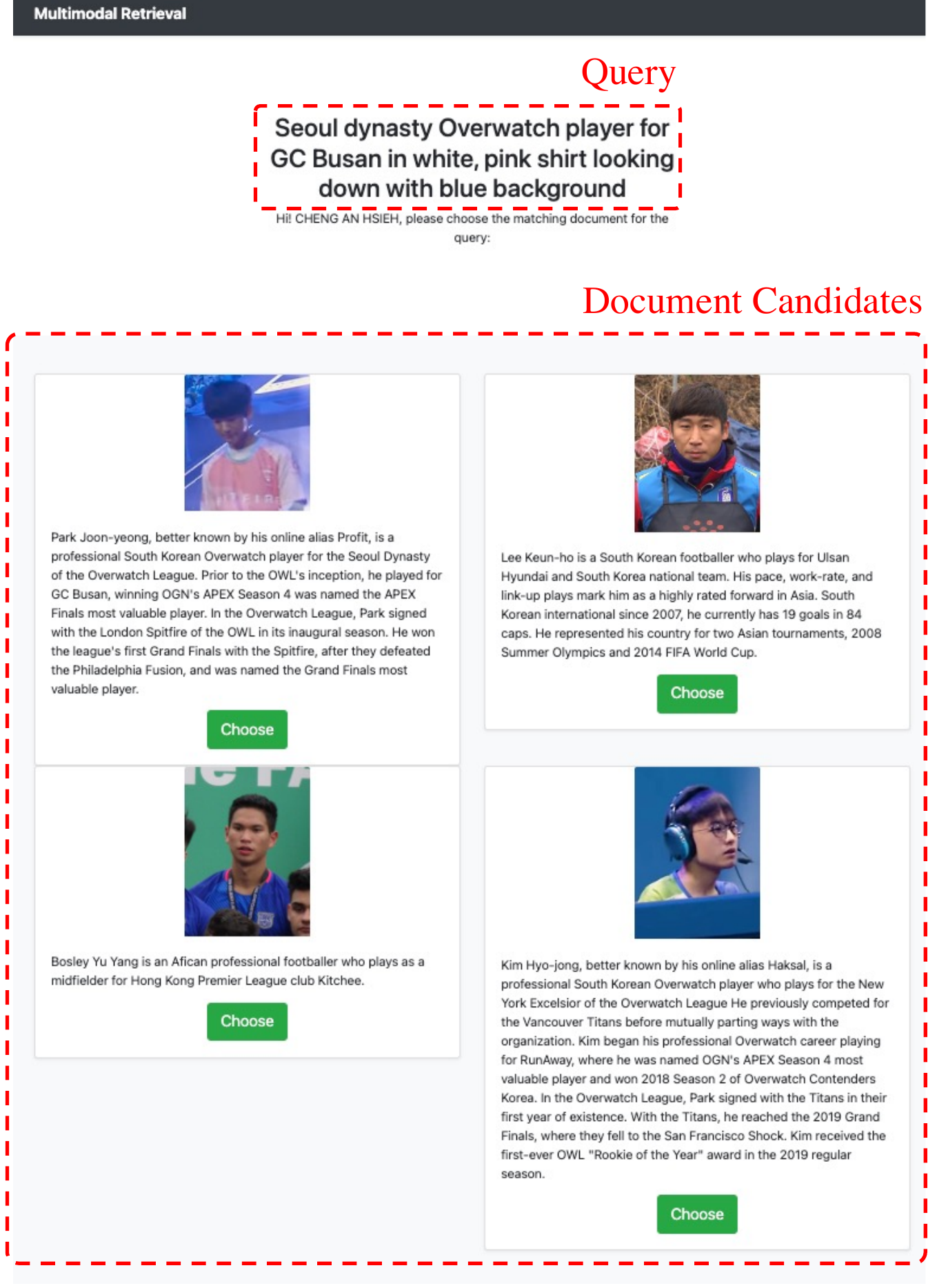}
    \caption{Template of human evaluation. }
    \label{fig:humaneval}
\end{figure}

\begin{figure}[H]
    \centering
    \includegraphics[width=\textwidth]{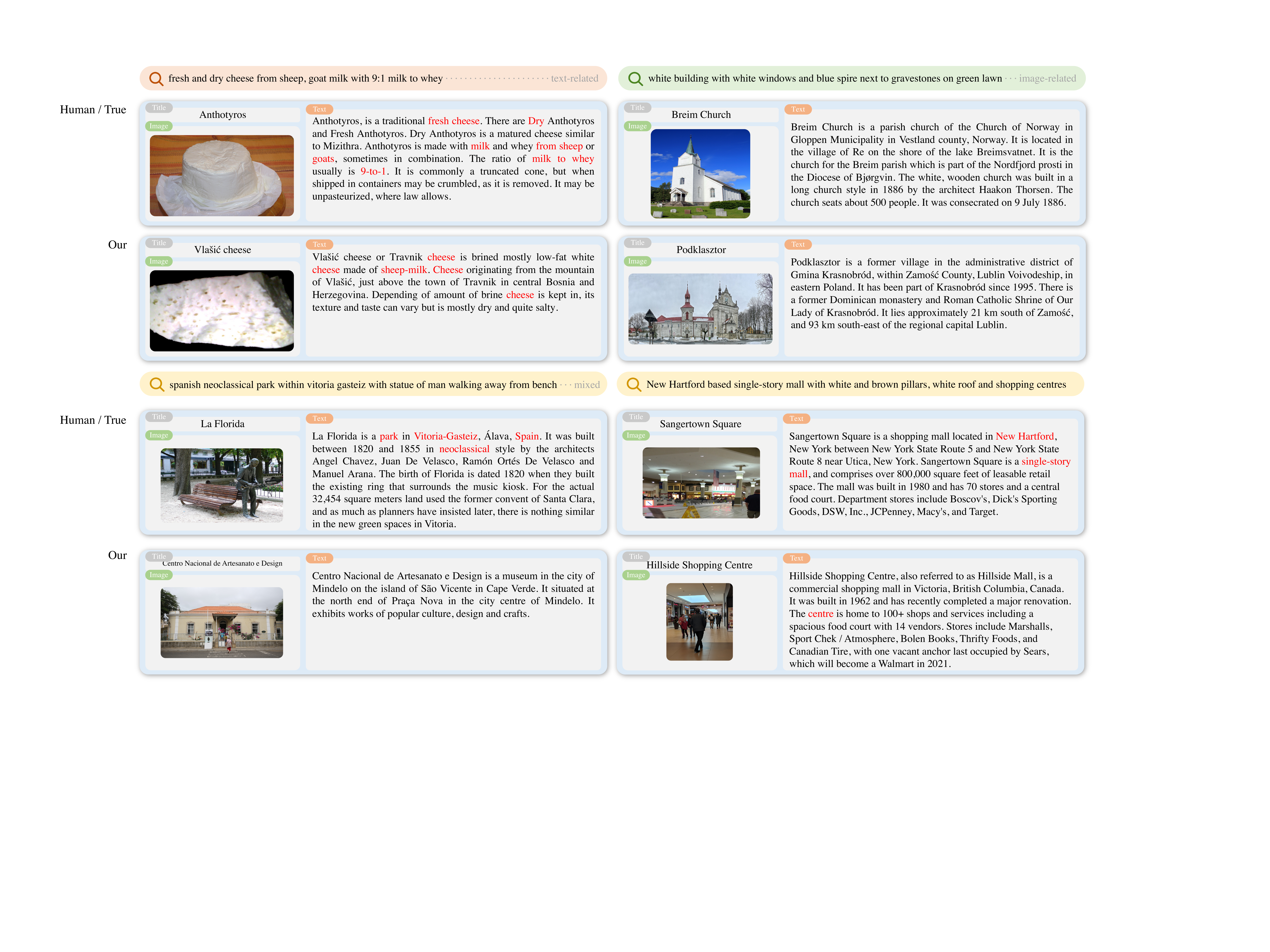}
    \vspace{-5pt}
    \caption{Failed examples of our model including text-related, image-related, and mixed queries.}
    \label{fig:fail}
\end{figure}

%  the words ``army''  and ``tank'' will attend corresponding parts on image while the words ``Australia'' and ``War'' will highlight ``Australian'' and ``fightning'' on texts
\subsection{Failed examples}
We create human evaluation by randomly sampling 50 examples from each type of task. As illustrated in Figure~\ref{fig:humaneval}, human annotators have to select the most relevant document from four candidates obtained from our MR model. Each question is answered by three workers. If more than half workers have the same answers and match the correct document, we consider humans can answer this question correctly; otherwise, humans fail it. 
With human evaluation, we can distinguish the performance difference between humans and our models in Figure~\ref{fig:fail}. For text-related queries, we can find our model is capable of obtaining the related document instead of the accurate one. However, humans can easily choose the right one by matching text keywords, such as ``9:1'' in the query and ``9-to-1'' in the document of the first example. For image-related queries, our model prefers to retrieve specific color words and ignores the remaining, such as the ``blue spire'' and ``green lawn'' of the second example. On the other hand, humans can perceive the details of images. For mixed queries, our model may pay attention to wrong words, such as the word ``walking'' and ``centres'' in the last two examples. On the contrary, humans can simultaneously recognize the correct document by matching text keywords and image context. 
\label{appendix-mistake}

\subsection{Compared to auto-generated and human-annotated queries}
\begin{wraptable}{r}{0.35\textwidth}
\centering
\vspace{-10pt}
\resizebox{0.35\textwidth}{!}{
\begin{tabular}{r|c}
\toprule
          & mixed query \\
          & MRR@10 \\
\midrule
Auto-generated & 32.6     \\
Human-annotated & 9.0     \\
\bottomrule
\end{tabular}}
\vspace{-3pt}
\caption{Retrieval performance of mixed queries from auto generation and human annotation.}
\label{tab:query}
\vspace{-10pt}
\end{wraptable}
Since our human-annotated queries have more diverse properties than auto-generated queries as shown in Table~\ref{tab:data-category}, we compare their performance on mixed queries with our pre-trained MR model (METER). The results are shown in Table~\ref{tab:query}, and we can find auto-generated queries outperform human-annotated queries because the annotated queries are more complex and difficult to learn. Therefore, we fine-tune our models on 1k annotated queries to adapt in human-annotated domain.
\label{appendix-queries}

\newpage
\section{Proposed Model Details}
\label{appendix-model}
With inputs document texts $D_T$ and document image $D_I$, we derive the document features $F_d$ from multimodal encoder $E_d$. Also, with input query 
texts $Q_T$, we obtain the query features $F_q$ from query encoder $E_q$. After, we take average of the size-variant features $F_d$ and $F_q$ to get a single fixed vector as document and query representations $R_d$ and $R_q$. 

\begin{equation}
    F_d = E_d(D_T, D_I) \quad \textrm{and} \quad F_q = E_q(Q_T)
\end{equation}
\begin{equation}
    R_d = Average(F_d) \quad \textrm{and} \quad R_q = Average(F_q)
\end{equation}

With document and query representations $R_d$ and $R_q$, we aim to close the distance between two vectors by contrastive learning. Therefore, we learn two projection functions $f_d$ and $f_q$ with fully-connected layers and L2-normalization to map their representations into the same space. We calculate the similarity by dot product for all document and query vector pairs in a training batch when treating matched pairs as positive while all other pairs as negative. The contrastive loss $L_{dqc}$ we minimize is in the following:

\begin{equation}
Sim(R_d, R_q) = f_d(R_d)^\top f_q(R_q)
\end{equation}

\begin{equation}
P^{i}_{d2q} = \frac{\exp(Sim(R^i_d, R^i_q) / \tau)}{\sum^{N}_{j=1}\exp(Sim(R^i_d, R^j_q) /\tau)},\ \ P^{i}_{q2d} = \frac{\exp(Sim(R^i_q, R^i_d)  / \tau)}{\sum^{N}_{j=1}\exp(Sim(R^i_q, R^j_d) /\tau)}
\end{equation}

\begin{equation}
L_{dqc} = - \frac{1}{B} \sum^{B}_{i}\frac{Y^{i}_{d2q}\log(P^{i}_{d2q}) + Y^{i}_{q2d}\log(P^{i}_{q2d})}{2}
\end{equation}

Here, we calculate the normalized softmax loss for both document-to-query and query-to-document classification. The loss is set up with batch size $B$, negative samples size $N(=B)$, and a learnable temperature parameter $\tau$ to scale the logits. For negative pairs of the document to query, $R^i_d$ and $R^j_q$ are the representations of the document in the $i$-th pair and query in the $j$-th pair, respectively. To more effectively organize in-batch negatives, we keep two queues \cite{li2021align} to store the most recent $K$ representations, helping enlarge the amounts of negative samples. The modified equation is to change negative sample size $N$ from batch size $B$ to queue length $K$.

In addition to contrastive loss, we obtain document-query matching loss to learn a fine-grained similarity of pair documents and queries. The matching loss is a binary cross-entropy loss to predict whether a pair of documents and queries are matched or mismatched. We build a 6-layer transformer-based classifier $C$ with input document features $F_d$ and query features $F_q$. Specifically, a special token (\emph{e.g.}, $[CLS]$) is inserted at the beginning of the input sequence, and it tries to learn a global cross-modal representation in transformers. After, a linear classifier is added to the $[CLS]$ token to predict a binary label. The whole matching loss is in the following: 

\begin{equation}
P^i_{dqm}(j=y^{i}_{dqm}) = \frac{\exp(C_j(F_d, F_q))}{\exp(C_0(F_d, F_q))+\exp(C_1(F_d, F_q))}
\end{equation}

\begin{equation}
L_{dqm} = - \frac{1}{B} \sum^B_iY^{i}_{dqm}\log(P^i_{dqm})
\end{equation}

Our full training objective is:
\begin{equation}
    L = L_{dqc}+L_{dqm}
\end{equation}
\newpage
\section{Dataset License}
\label{appendix-license}
Our dataset is under the Creative Commons Attribution Share Alike 4.0 (CC BY-SA 4.0) license.
\section{Maintenance}
\label{appendix-maintenance}
We believe that Mr. Right will assist researchers in building robust multimodal retrieval models and improve the current retrieval systems. We are willing to maintain Mr. Right. If researchers have any problems, they can create an issue from our repository. We also welcome any methods to perform on our benchmark. We bear all responsibility for violations of rights related to Mr. Right.

\end{document}